\documentclass[preprint,12pt,authoryear]{elsarticle}

\usepackage{amssymb}
\usepackage{graphicx}

\journal{Planetary and Space Science}
\begin{document}

\begin{frontmatter}



\title{IAU MDC Meteor Orbits Database - A Sample of Radio-Meteor Data from
the Hissar Observatory}

\author[mn]{M. Narziev}
\author[rpc]{\underline{R. P. Chebotarev}}
\author[tjj]{T. J. Jopek}
\author[ln]{L. Neslu\v{s}an}
\author[vp]{V. Porub\v{c}an}
\author[ln]{J. Svore\v{n}}
\author[mn]{H. F. Khujanazarov}
\author[rpc]{\underline{R. Sh. Bibarsov}}
\author[rpc]{\underline{Sh. N. Irkaeva}}
\author[rpc]{\underline{Sh. O. Isomutdinov}}
\author[rpc]{\underline{V. N. Kolmakov}}
\author[rpc]{\underline{G. A. Polushkin}}
\author[rpc]{\underline{V. N. Sidorin}}

\address[mn]{Institute of Astrophysics of Academy of Sciences of\\
           the Republic of Tajikistan, Tajikistan}
\address[rpc]{Institute of Astrophysics of Academy of Sciences of\\
           the Republic of Tajikistan, Tajikistan\footnote{Dr. Roman
 Petrovich Chebotarev made a lot of work to prepare the catalog
 presented in this paper. He was the head of the radar laboratory and
 leader of the team creating the radio complex MIR-2, which was
 essential to collect the Hissar database. Unfortunately, he and
 another workers collecting and processing the data (Drs. R. Sh.
 Birbasov, Sh. N. Irkaeva, Sh. O. Isomutdinov, V. N. Kolmakov, G. A.
 Polushkin, and V. N. Sidorin) passed away before the work was finished
 and published.}}
%           \email{mirhusseyn\underline{~}narzi@mail.ru}
%                  \and
% $[$Dr. Roman Petrovich
% Chebotarev made a lot of work to prepare the catalog presented in
% this paper. Unfortunately, he passed away before the work was finshed.
% He was the head of the radar laboratory and leader of the team
% creating the radio complex MIR-2, which was essential to collect the
% Hissar database.$]$
%                  \and
\address[tjj]{Astronomical Observatory Institute, Faculty of Physics,
           Adam Mickiewicz University, \\
           ul. Sloneczna 36, 60-286 Pozna\'{n}, Poland}
%           \email{jopek@amu.edu.pl}
%                  \and
\address[ln]{Astronomical Institute, Slovak Academy of Sciences, \\
           05960 Tatransk\'{a} Lomnica, Slovakia}
%           \email{ne@ta3.sk, astrsven@ta3.sk}           %  \\
%                  \and
\address[vp]{Astronomical Institute, Slovak Academy of Sciences, \\
           05960 Tatransk\'{a} Lomnica, Slovakia \& \\ 
           Faculty of Mathematics, Physics and Informatics,
           Comenius University, \\
           84248 Bratislava, Slovakia}
%           \email{porubcan@fmph.uniba.sk}
%           %            \emph{Present address:} of F. Author  %  if needed
% \address[js]{Astronomical Institute, Slovak Academy of Sciences, \\
%           05960 Tatransk\'{a} Lomnica, Slovakia}

\begin{abstract}
We announce an upgrade of the IAU MDC photographic and video meteor
orbits database. A sample of 8916 radio-meteor data determined by the
radar observations at the Hissar (Gissar) Astronomical Observatory,
Dushanbe, Tajikistan, are added to the database. Along with the radiant
coordinates, velocities, and orbital elements the Hissar radio-meteor
sample contains the heights, linear electron densities, stellar
magnitudes, and masses of the meteoroids. The new $2018$ version of
the IAU MDC database contains $4873$ photographic, $110521$ video, and
$8916$ radio meteor records. The data are freely available on the
website at the address {\em https://www.astro.sk/\~\,ne/IAUMDC/PhVR2018/}.
\end{abstract}


%Research highlights
\begin{highlights}
\item Radio-meteor data from the Hissar Astronomical Observatory  released
\item Method of radio-meteor observations described
\item IAU MDC database extended
\item Sample of radio-meteor orbits freely available
\end{highlights}


\end{frontmatter}



% 1.
\section{The orbital database of the IAU Meteor Data Center}
\label{sect1}

Our knowledge of the structure of meteoroid streams as well as an abundance
of the sporadic meteoroids has been derived, mostly, from observations of
the meteor phenomena in the Earth's atmosphere. Several research teams
systematically or during observational campaigns recorded meteors using
different observational techniques: traditional and digital photography,
radar, video and TV equipment.  Obtained results were published in a
printed form or in the way of a local database. Since the eighties of
the previous century \citep{1987PAICz..67..201L} the IAU Meteor
Data Center (MDC) collects the orbital and geophysical parameters of the
individual meteoroids. These data are freely accessible on the dedicated
website posted at the server maintained by the Astronomical Institute of
Slovak Academy of Science in Tatransk\'{a} Lomnica.
The last version of the IAU MDC database issued in 2016 contains the
collection of $4873$ photographic records
\citep{2003EM&P...93..249L, 2014EM&P..111..105N, 2016CBET...4255N} and
the new video-based catalog of $110521$ meteors observed by the Cameras
for All-sky Meteor Surveillance  system (CAMS),
\citep{2011pimo.conf...28G, 2011JIMO...39...93J, 2011Icar..216...40J,
2016Icar..266..371J, 2016Icar..266..331J, 2016Icar..266..355J,
2016Icar..266..384J}.

   %
% FIG. 1
\begin{figure}[t]
\vbox{
\hbox{
\centerline{
\includegraphics[width=5.8cm]{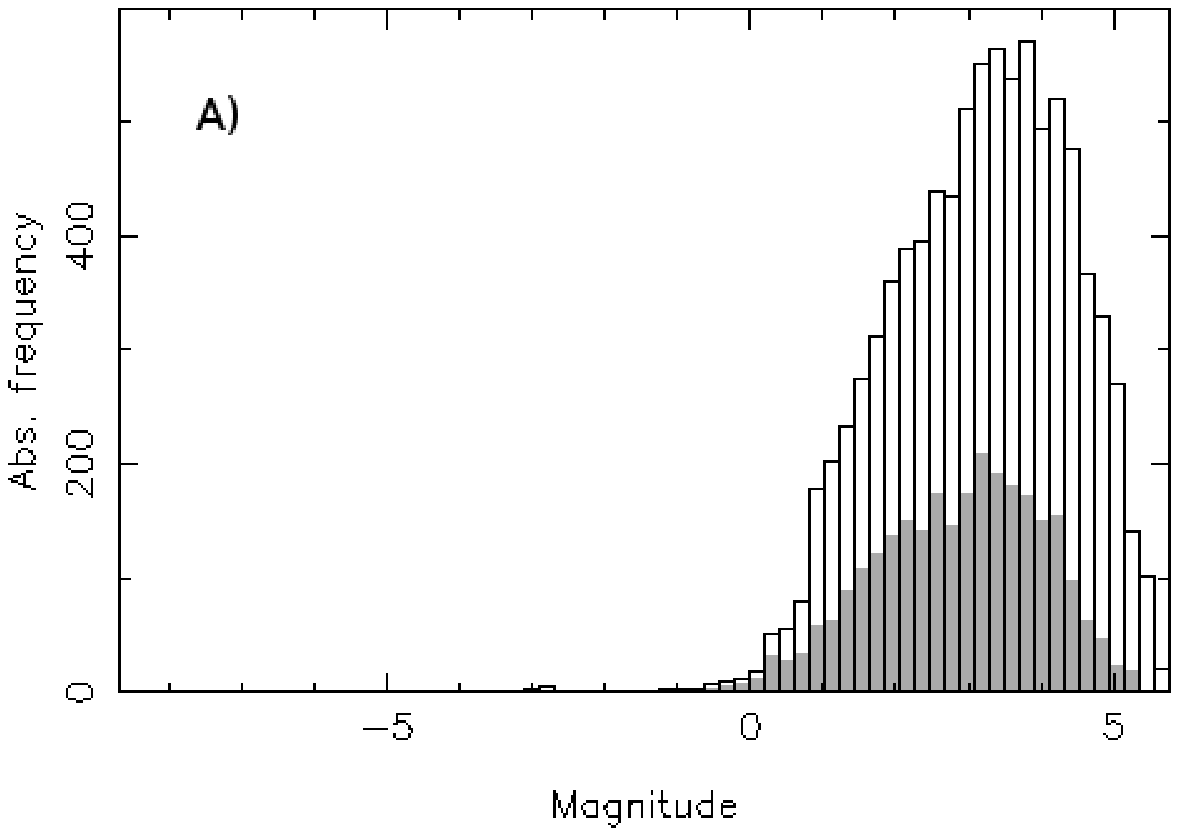}  \hspace{0.1cm} 
\includegraphics[width=5.8cm]{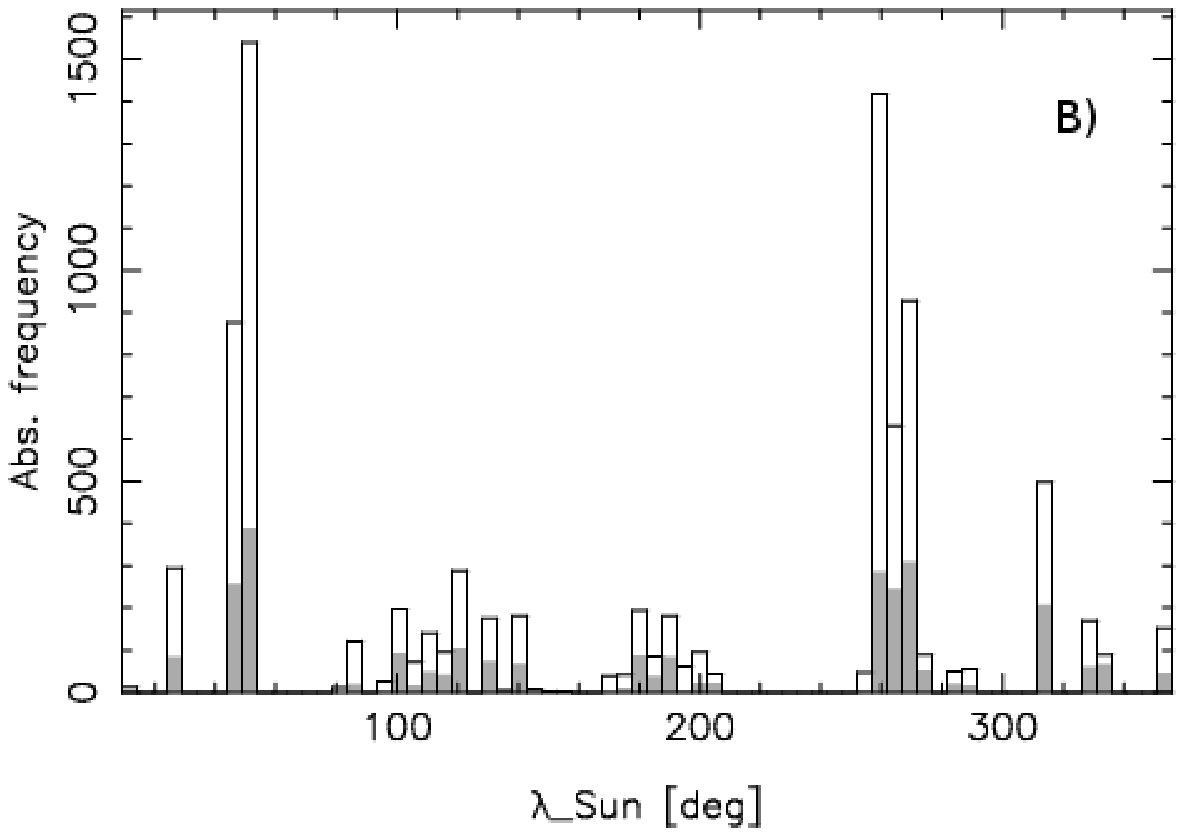} 
} % cent
} %1
\hbox{
\centerline{
\includegraphics[width=5.8cm]{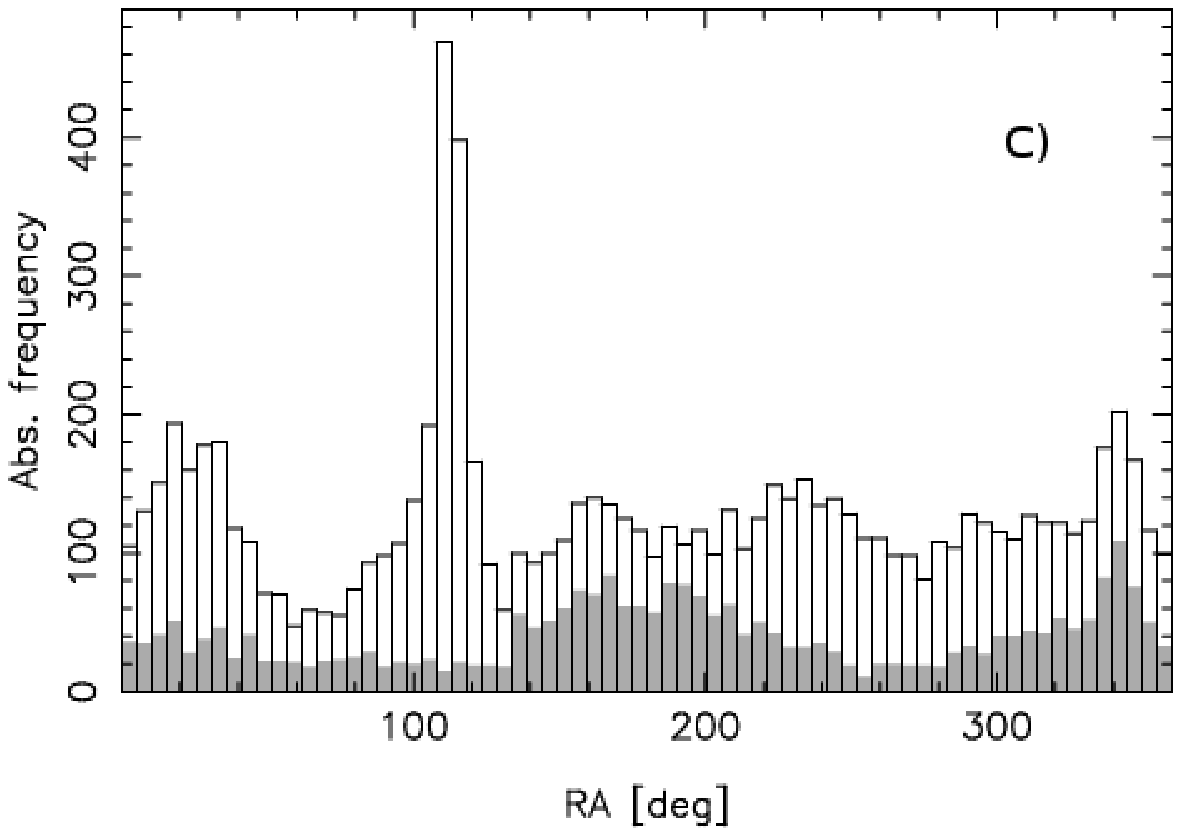}  \hspace{0.1cm} 
\includegraphics[width=5.8cm]{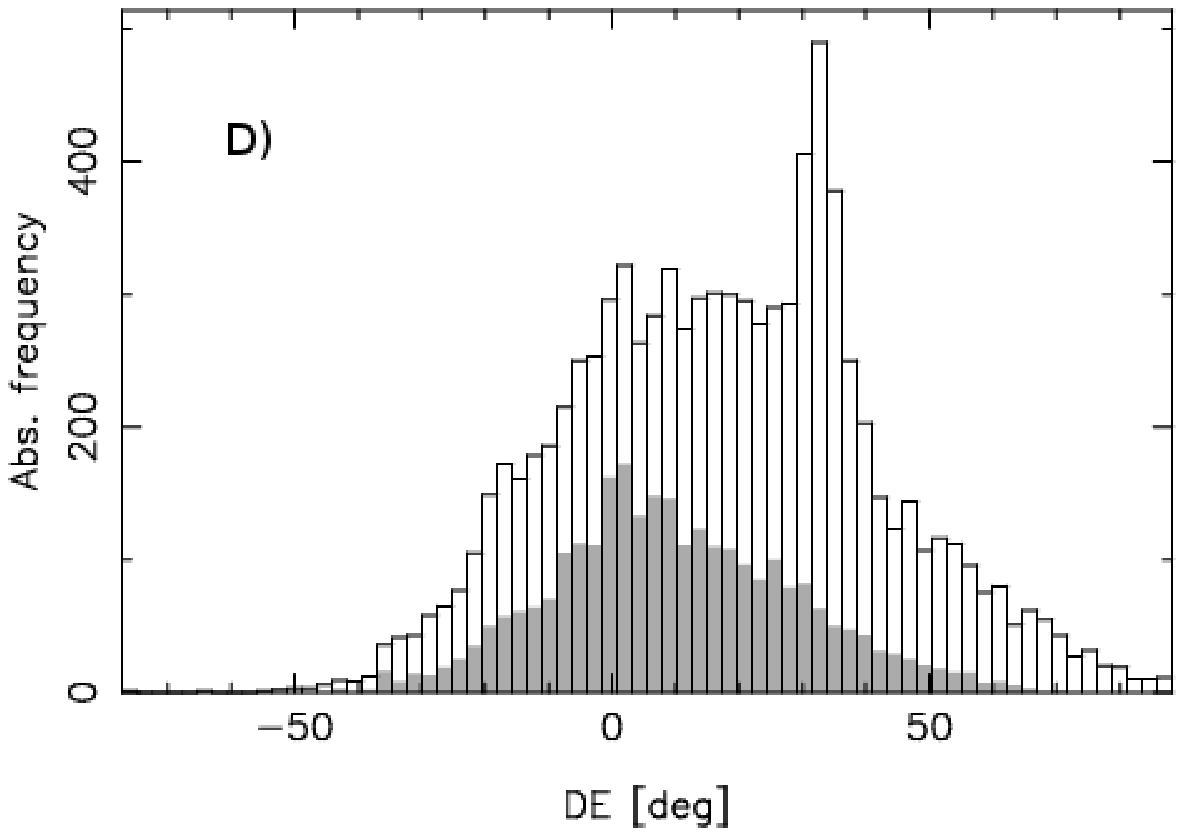} 
} % cent
} %2
\hbox{
\centerline{
\includegraphics[width=5.8cm]{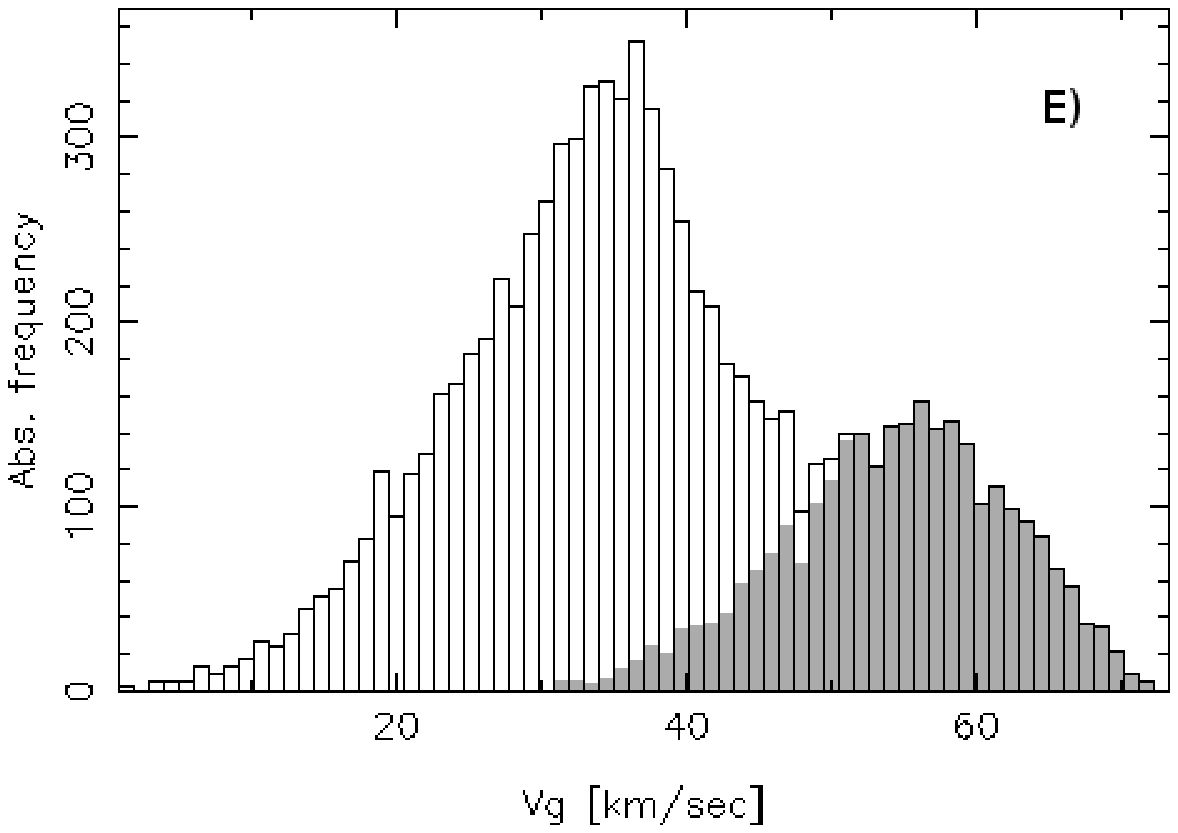}  \hspace{0.1cm} 
\includegraphics[width=5.8cm]{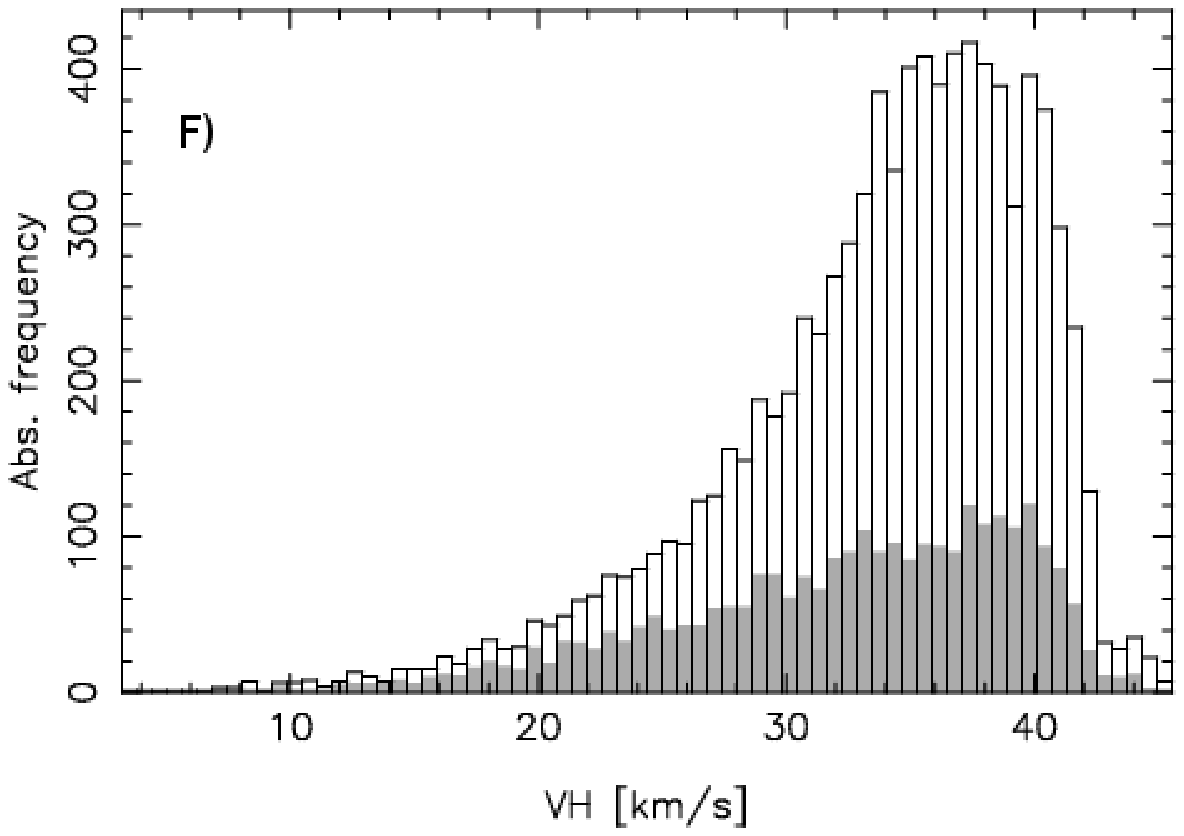} 
} % cent
} %3
}
\caption[f1]{The histograms of selected parameters of $8916$ radio-meteors
observed in the Hissar station. The shadowed bins refer to the retrograde
orbits.       
A) --- the values of the meteor radar magnitude; mostly in the range of
$0$--$5.5^m$ with maximum at $3.3^m$. 
B) --- the ecliptic longitude of the Sun at the meteor apparition.
The bulk of meteors in the pick on the right were observed during
the December activity of Geminids meteor shower. C) D) --- the right
ascension and declination of the meteor geocentric radiants. The prominent
picks correspond to the Geminids. E) F) --- the geocentric and heliocentric
velocities of the meteoroids. The distributions have different properties,
the first has two maxima, the latter only one, see text. 
}
\label{FIG1}
\end{figure}
The photographic and video observations provide us only with the information
about the night-time\footnote{The commonly accepted definition of the
day-time or night-time meteor shower does not exist. In practice some
researches use the rule--- the elongation of the radiant of a day-time
shower from the Sun is less than $20$--$30$ deg
\citep{2014pim4.conf...93R}.} meteors: the shower and the sporadic
components. Admittedly, some video databases
\citep{2009JIMO...37...55S, 2016JIMO...44...42S} contain a small
fraction of the day-time meteors, but the day-time  showers have mostly
been studied on the basis of the radio-meteor data. The reason is simple,
the radio-meteor observations do not depend on any specific atmospheric
or day-night conditions.
Therefore we find it valuable to incorporate to the IAU MDC collection
a sample of $8916$ radio-meteor data gained at the
Hissar\footnote{Because of the transcription of Russian texts to English,
name ``Hissar'' has often be written as ``Gissar''.} Astronomical
Observatory in Tajikistan.

The dates of observations of $8916$ radio-meteors span the synoptic-year
interval from  December $1968$ until  December $1969$. Such rather small
sample do not enable a complex study of meteor showers, but it can serve
for various comparisons between the radar, video, and photographic
techniques. In this paper, we introduce the Hissar radio-meteor data which
are available on the IAU MDC website
({\em https://www.astro.sk/\~\,ne/IAUMDC/PhVR2018/}).
%

% 2.
\section{Apparatus and data processing}
The Hissar radio-meteor station of the Institute of Astrophysics,
Academy of Sciences of Tajikistan is located in Dushanbe; its geographic
coordinates are: $\phi$=$38.6^o\,$N, $\lambda$=$68.8^o\,$E. The meteors
were observed by the radar complex MIR-$2$ system (the meteor pulse radar
of the second generation) which was constructed in $1964$-$1968$. Its
detailed structure, principles of operation, and technical characteristics
are described in \citep{1970BIAst..55...24C, 1970BIAst..55...34T}.

The MIR-$2$ system consisted of a transmitter working at the wavelength
$8\,$$[$m$]$ ($37.4\,$$[$MHz$]$) with the output pulse power of
$65\,$$[$kW$]$ and five receiving antennas. Four of them were located at
$3.9$ to $4.0\,$$[$km$]$ from the central receiver. The bandwidth of the
primary and remote receivers was $600\,$$[$kHz$]$. 
%Such positions of the receivers allowed to record a sketch of the ionization curve with the length to $3.75\,$[km]. 
The receivers sensitivity threshold was equal to $8\cdot 10^{-14}\,$$[$W$]$.
For further processing only those meteors were used which were recorded by
all receivers and had at least first maximum in the amplitude-time
characteristics (ATC), and which were not substantially distorted by the
noise. 

The results of the radar observations displayed on the cathode-ray tubes
were recorded photographically. For recorded meteors it was found: the
date and time of the meteor appearance, the distance to the meteor trail
from the main central antenna, the distance to the meteor trail from each
of the four receiving antennas.
For each receiving channel of the amplitude-time characteristics (ATC),
the positions of the beginning and the first four maxima were measured.
Using these data and two methods described in \citep{1970KomMe..19...46C}
and \citep{1976KomMe..24...19C} the angular coordinates of the azimuth
and the zenith distance of the mirror point on the meteor trail and the
velocity of the meteoroid were determined.
If the value of the measured radio-echo duration for the central point
enables to determine the value of the linear electron density, then the
height of meteor trail, together with the radio magnitude and mass of
a meteoroid were also determined, simultaneously with the radiant
coordinates and velocity.

If the distances between the central and four remote receiving points
are around $4\,$[km], then using the bearing-time method proposed by
\citet{1976KomMe..24...19C}, the accuracy of the MIR-$2$ system
estimated by the root-mean-square errors (RMSE) is as follows:
$\sigma_d$=$\pm 20\,$[m] for the distances,
$\sigma_t$=$\pm 1\,$[millisecond] for time measurements. In case of the
long-range version of the bearing-time method the RMSE of the azimuth of
the meteor radiant $\sigma_{AR}$=$\pm 1.8^{o}/\sin Z_{R}$, and the zenith
distance $\sigma_{ZR}$=$\pm 1.2^{o}$, where $Z_R$ is the zenith distance
of the radiant. The relative error of the meteor velocity equals to $3\%$.
The RMSEs of the azimuth and zenith distance of the reflecting point are
$\sigma_{A}$=$\pm 0.9\sin Z$ and $\sigma_{Z}$=$\pm 0.9\cos Z$
respectively, where $Z$ is the zenith distance of the reflecting point.
The RMSE of the height of the reflecting point relative to the main
remote station equals to $\sigma_h$=$\pm 2.0\,$[km].

It should be noted that in case of the bearing-time method one can use
the amplitude-time pictures of inferior quality with $1$-$2$ extremes
only, instead of $3$-$4$ which are necessary for the pulse-diffraction
method. This approximately doubles the sensitivity of the method (based
on the number of measured radiants and velocities), but increases the
error $\sigma_{A}$ by the factor of $1.5$. The speed of the meteor in
the bearing-time method was determined both by the time of flight of the
individual sections of the meteor path and using the diffraction pattern.
Consistency of these velocities allowed to estimate the accuracy of the
measurements. The final measured speed was either the average of the
two, or it equalled to the velocity with a smaller measurement error.
%

%3.
\section{Hissar radio-meteor data sample}
   The Hissar $8916$ meteors were observed within the period from
December 12, 1968 until December 24, 1969. We note that these data were
published by \citet{2019Donish} before they were included to the IAU MDC
database.

   Most of their radio magnitudes spans the interval $0$--$5.5^m$ (see
Figure~\ref{FIG1}A) with maximum at $3.3^m$. However isolated cases of
one bright bolide ($-8.6^m$) and bright meteoroids were recorded. The
bulk masses of the meteoroids falls into interval $10^{-4}$--\,$1$ [g].
   The distribution of ecliptic longitudes of the Sun at the meteor
apparitions, as shown in Figure~\ref{FIG1}B, is not a uniform one. A few
gaps and two  distinct picks are seen. The December pick certainly
corresponds to the Geminids shower (\#4/GEM)\footnote{The shower
designations \#4/GEM fulfills the meteor shower nomenclature rules
delineated in, e.g., \citep{2011msss.conf....7J,2017P&SS..143....3J}.}.
   The second pick observed in May is not related with a prominent meteor
shower. Instead, in this month, in most cases, the sporadic meteors were
observed. This is clearly seen in Figure~\ref{FIG1}C and ~\ref{FIG1}D --
where only one pick corresponding to Geminids geocentric radiants is seen.
Using the break-point method
\citep{1995EM&P...68..427N, 2013EM&P..110...41N}, we found that in the
Hissar sample the relatively well-defined Geminid stream would consist of
$\sim$800 meteoroids.

% FIG. 2
\begin{figure}[t]
\centerline{
\includegraphics[width=7.2cm, angle=-90]{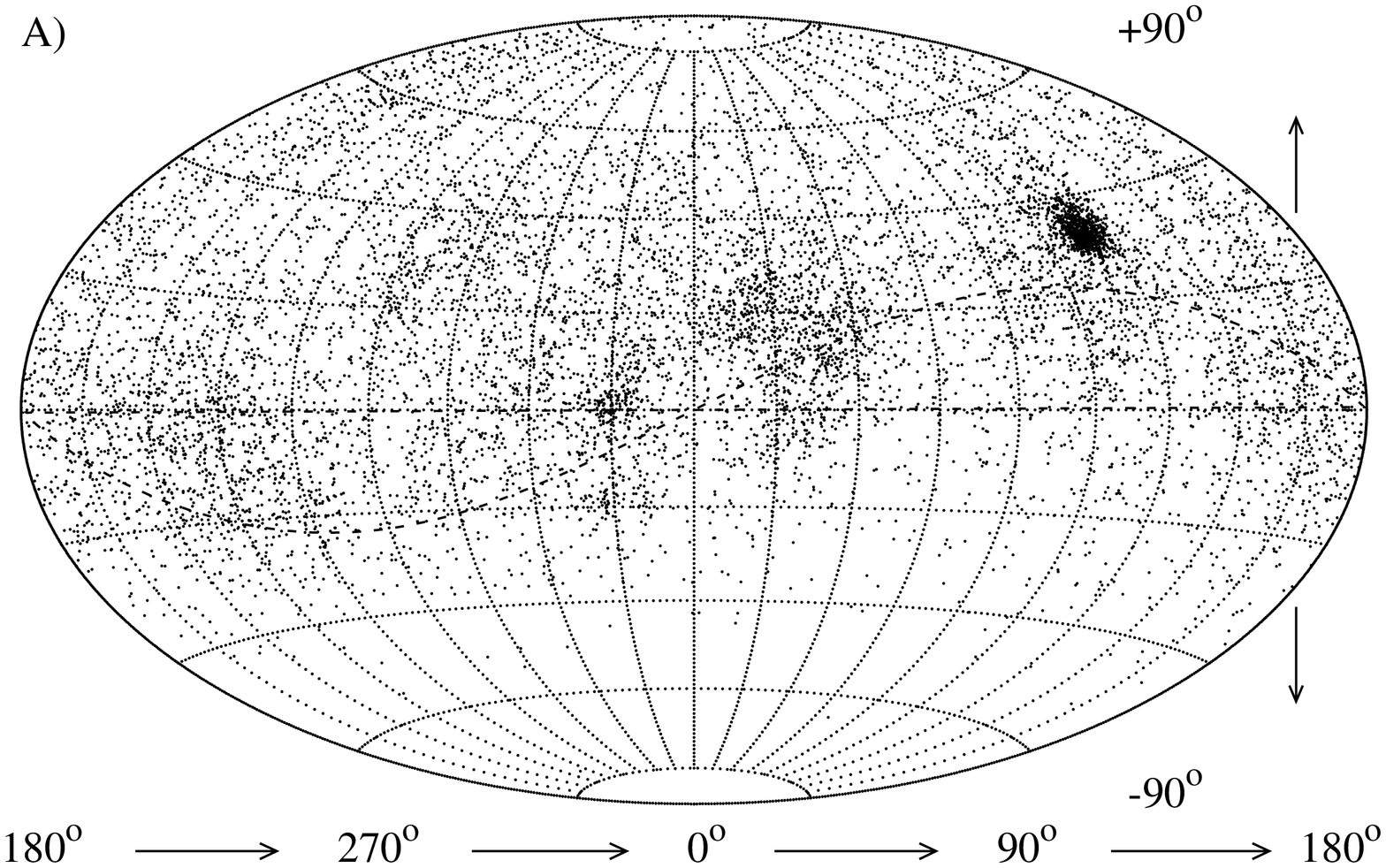}
}
\vspace{-0.5cm}
\centerline{
\includegraphics[width=7.2cm, angle=-90]{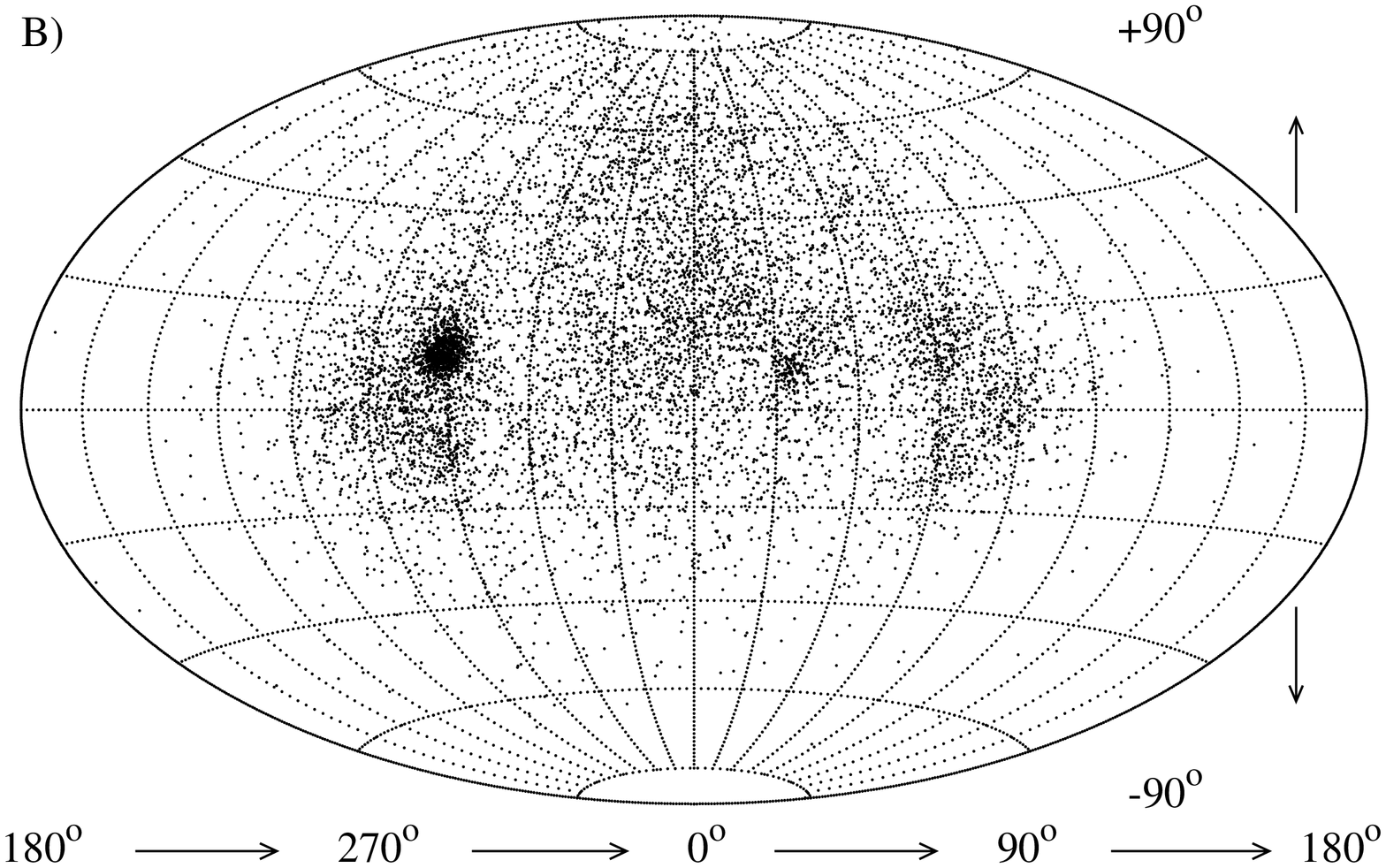} 
} %
\vspace{-0.5cm}
\caption[fha]{The Hammer-Aitoff diagram of the geocentric radiants of
$8916$ Hissar meteors. In the diagram, the sky is seen from a position
situated outwardly the celestial sphere. In plot A), the equatorial
coordinates are used, the sinusoid-like dashed curve represents the
ecliptic. In plot B), the same radiants are plotted in the ecliptic
coordinates with the ecliptic longitude shifted in such a way that the
apex of the Earth's motion around the Sun is in the origin of the
coordinate frame. Due to the geographical location of the Hissar
station, many sporadic meteors with southern radiants were observed. On
the right in plot A), the very compact concentration of the Geminids is
clearly seen. Two other widespread clusters nearby the ecliptic are
located around the center of the upper diagram.
}
\label{hara}
\end{figure}

   The histogram of the geocentric velocity is plotted in
Figure~\ref{FIG1}E and its characteristic bimodal distribution is
apparent. The bimodal nature of this distribution arise from that
the values of geocentric velocities are superposition of the two
components, the meteoroid and the Earth heliocentric orbital
velocities. Figure~\ref{FIG1}F illustrates the distribution of the
meteoroid heliocentric velocity. In the right tail  of this distribution
we noted that $148$ meteoroids moved on the hyperbolic heliocentric
orbits. Most of them (possibly all) are not really hyperbolic and they
occurred due to measurement uncertainties, particularly due to the
uncertainty of the geocentric velocity.
   However, considering the recent discovery of 1I/`Oumuamua, a mildly
active comet and the first interstellar small body recorded inside the
solar system, we shouldn't neglect the existence of the interstellar
meteoroids. 
   The Hammer-Aitoff projection of the geocentric radiants of $8916$
radio-meteors is illustrated in Figure~\ref{hara}. Most radiants are
located on the northern hemisphere. Except of Geminids and some not yet
identified widespread clusters at the center of diagram 2A, the radiants
located above the cellestial equator are distributed almost uniformly. In
Figure~\ref{hara}B the ecliptic Earth apex centered reference frame was
used. The helion and antihelion concentrations are clearly seen.
%

% 4.
\section{IAU MDC database: the list of parameters and the data format}
The 2016 version of the MDC orbital database includes $41$ photographic
catalogs \citep{2014EM&P..111..105N} and one video meteor catalog
\citep{2016CBET...4255N}. 

Each individual catalog incorporates some amount of the data records and
each data record contains the values of several dozen of compulsory and
supplementary parameters. The eleven compulsory parameters are: the
identification code, the date of meteor fall, the orbital elements
(perihelion distance, eccentricity, inclination, argument of perihelion,
and longitude of ascending node), the geocentric radiant coordinates
(right ascension and declination), and the geocentric and heliocentric
velocities. 

The supplementary parameters can be different depending on the
observational technique used. The unification of all parameters of all
individual catalogs represents the maximal meteor data record stored in
the database. In the database version $2016$, the maximal meteor data
record consisted of 31 parameters. It was adjusted to the photographic
and video meteor data.

Incorporation of the radio-data required adding another three parameters:
the stellar magnitude of the radio-meteor, the height of the central point
of the meteor trail, and the linear electron density of the central point
of the meteor trail (a common logarithm of this quantity). 
The complete list of $34$ meteor parameters of the current 2018 version
of the MDC database is given in Table~\ref{TAB1}.
%
% TAB. 1
\begin{table}
\caption{The list of parameters included in the new 2018 version of the
IAU MDC orbital database. No.P. is the serial number of the parameter; C.P. is the code of the parameter. For the angular parameters Equinox 2000.0 is obligatory.
The asterisks mark the parameters which are regarded as compulsory.}
\label{TAB1}
\begin{tabular}{rll}
\hline \hline \noalign{\smallskip}
 No.P. & C.P. & Explanation \\
\hline
$*$  1 & $\#$IC: & IAU MDC identification code \\
     2 & ANo: & unique number/code assigned to the meteor by  the author \\
$*$  3 & Yr~: & year of the meteor detection \\
$*$  4 & Mn~: & month of the meteor detection \\
$*$  5 & Day: & day and fraction of day of the detection in the UTC time\\
       &      & ~~~~~scale\\
  6 & LS~: & solar longitude corresponding to the date of the meteor\\
    &      & ~~~~~detection $[$deg$]$ \\
  7 & mv~: & maximum photographic brightness of meteor $[$magnitude$]$ \\
  8 & HB~: & height of the beginning of meteor trail $[$km$]$ \\
  9 & HM~: & height of the maximum brightness $[$km$]$ \\
 10 & HE~: & height of the end of meteor trail $[$km$]$ \\
$*$ 11 & RA~: & right ascension of the geocentric radiant $[$deg$]$ \\
$*$ 12 & DEC: & declination of the geocentric radiant $[$deg$]$ \\
 13 & Vi~: & extra-atmospheric velocity $[$km$\,$s$^{-1}]$ \\
$*$ 14 & Vg~: & geocentric velocity $[$km$\,$s$^{-1}]$ \\
$*$ 15 & Vh~: & heliocentric velocity $[$km$\,$s$^{-1}]$ \\
 16 & cZ~: & cosine of the zenith distance of geocentric radiant \\
 17 & Qm~: & quality code \\
$*$ 18 & q~~: & perihelion distance $[$AU$]$ \\
$*$ 19 & e~~: & numerical eccentricity of the orbit \\
 20 & 1/a: & reciprocal semi-major axis $[$AU$^{-1}]$ \\
 21 & a~~: & semi-major axis $[$AU$]$ \\
 22 & Q~~: & aphelion distance $[$AU$]$ \\
$*$ 23 & i~~: & inclination of the orbit to the ecliptic $[$deg$]$ \\
$*$ 24 & arg: & argument of perihelion $[$deg$]$ \\
$*$ 25 & nod: & longitude of ascending node $[$deg$]$ \\
 26 & pi~: & longitude of perihelion $[$deg$]$ \\
 27 & Sh~: & the IAU meteor shower code \\
 28 & Mas: & pre-atmospheric photometric mass $[$g$]$ \\
 29 & lgM: & common logarithm of the mass \\
 30 & cor: &  remark on correction (if appears) \\
\noalign{\smallskip} \hline
\end{tabular}
\end{table}

\addtocounter{table}{-1}
\begin{table}
\caption{$-$ continuation.}
\begin{tabular}{rll}
\hline \noalign{\smallskip}
 No.P. & C.P. & Explanation \\
\hline
 31 & crh: &  remark on extreme hyperbola \\
 32 & mr~: &  magnitude (stellar) of radio meteor \\
 33 & Hrf: &  the height of the central point of the meteor trail $[$km$]$ \\
 34 & LpA: &  common logarithm of linear electron density of \\
    &      &  the central point of meteor trail, $p_{\alpha}$
              $[$electron$\,$cm$^{-1}$$]$ \\
\noalign{\smallskip} \hline \hline
\end{tabular}
\end{table}

The data in the MDC database version 2018 are stored in the standard and
single-line (reduced-data) formats. The data formats were introduced in
version 2013 \citep{2014EM&P..111..105N} and since then, the formats
are fixed and used in each subsequent version, version 2018 including.

Since the database version 2016 \citep{2016CBET...4255N}, the IAU MDC
provides the data in the Excel sheet format. In version 2018, this format
is unchanged in the case of the photographic and video CAMS data. We were,
however, forced to establish a new form of the Excel sheet for the Hissar
radio-data because of the different set of the supplementary parameters
submitted with this catalog. There are 23 columns with 23 parameters
arragned in order:\\

\noindent
$\#$IC, Yr, Mn, Day, Hrf, mr, RA, DEC, Vi, Vg, Vh, cZ, q, e, 1/a, a, Q, i, arg, nod, pi, LpA, lgM.\\

\noindent
The codes are explained in Table~\ref{TAB1}. Information about a given
meteor is written in single line.

As already mentioned, the 2018 version of the IAU MDC orbital database
can be freely downloaded from the website:\\
{\em https://www.astro.sk/\~\,ne/IAUMDC/PhVR2018/}. In each available
format, all data can be downloaded as a single compressed ZIP file or
each component catalog can be downloaded separately as a plain file.

We recapitulate that the upgraded $2018$ IAU MDC orbital database contains
$4873$ photographic, $110\,521$ video, and $8916$ radio-meteor records.
Also, we remind that detailed descriptions of the data sources, listed
parameters, formats, and corrections introduced to the submitted data are
given in the database documentation which is also included in the ZIP
archives or can be downloaded separately as the PDF file.

The IAU MDC orbital database is maintained on an unpaid voluntary basis.
Several colleagues have asked us --- what reference to the orbital
database they should include in a published paper? The correct references
are: 
\begin{itemize}
 \item
 Porub\v{c}an, V., Svore\v{n}, J., Neslu\v{s}an, L., Schunov\'{a}, E., 2011,
 The Updated IAU MDC Catalogue of Photographic Meteor Orbits. Proc. of the
 Meteoroids Conference held in Breckenridge, Colorado, USA, May 24--28,
 2010, Meteoroids: The Smallest Solar System Bodies. NASA/CP--2011-216469,
 Cooke, W. J., Moser, D. E., Hardin, B. F., and Janches, D. (eds),
 pp 338-341.
 \item  
 Neslu\v{s}an, L., Porub\v{c}an, V., Svore{\v n}, J., 2014, IAU MDC
 Photographic Meteor Orbits Database: Version 2013, Earth Moon and Planets
 111, pp 105-114. 
 \item
 Neslu{\v s}an, L.,  Jenniskens P., Porub{\v c}an, V., Svore{\v n}, J.,
 2016,  Central Bureau Electronic Telegram No. 4255. 
 \item and also this paper.
\end{itemize}

\indent \\

%    We will appreciate all users of the IAU MDC Photographic Meteor
% Orbits Database if they refer the database by citeing Porub\v{c}an et al.
% (2011) and this paper.

\small
\noindent
{\bf Acknowledgements.}
This work has been supported, in part, by the VEGA - the Slovak Grant
Agency for Science, grants Nos. 2/0023/18 (J. Svore\v{n}) and 2/0037/18
(L. Neslu\v{s}an), and by the Slovak Research and Development Agency under
contract APVV-16-0148 (V. Porub\v{c}an).

T.J. Jopek contribution was supported by 2016/21/B/ST9/01479 project,
founded by the National Science Centre in Poland. 

This research has made use of NASA's Astrophysics Data System Bibliographic
Services. 

\normalsize
% \bibliographystyle{spbasic}
% \biboptions{authoryear}    % basic style, author-year citations
%\bibliographystyle{spmpsci} %\biboptions{authoryear}      % mathematics and physical sciences
%\bibliographystyle{spphys}  %\biboptions{authoryear}     % APS-like style for physics
%\bibliographystyle{apalike.bst}
%\bibliography{bibliography.bib}   % name your BibTeX data base

%\end{document}

\end{document}